\documentclass[SIMONE,PDF]{cej-simone2} 
\usepackage{layout}
\usepackage{amsfonts}
\usepackage{amsmath}
\usepackage{multirow}
\usepackage{fancyref}
\usepackage{natbib}
\usepackage{hyperref}
\usepackage[english]{babel}
\def\ajpsi   {\ensuremath{\alpha_s(\mjpsi)}}
\def\B    {\ensuremath{\mathcal{B}}}

\def\ug       {\ensuremath{\!\!\!&=&\!\!\!}}

\def\ie       {i.e.}

\def\hh       {\hspace{10 mm}}

\def\ee       {\ensuremath{e^+e^-}}

\def\bfr      {\begin{flushright}}
\def\efr      {\end{flushright}}
\def\bm       {\begin{minipage}}
\def\em       {\end{minipage}}
\def\jpsi     {\ensuremath{J/\psi}}
\def\psii     {\ensuremath{\psi(2S)}}
\def\pipi     {\ensuremath{\pi^+\pi^-}}
\def\ee     {\ensuremath{e^+e^-}}
\def\pip     {\ensuremath{\pi^+}}
\def\pim     {\ensuremath{\pi^-}}
\def\piz     {\ensuremath{\pi^0}}
\def\opiz     {\ensuremath{\omega\pi^0}}

\def\gp     {$G$-parity}

\def\mjpsi     {\ensuremath{M_{J/\psi}}}
\def\gjpsi     {\ensuremath{\Gamma_{J/\psi}}}

\def\EM    {electromagnetic}
\def\aggg    {\ensuremath{\mathcal{A}_{3g}}}
\def\agg    {\ensuremath{\mathcal{A}_{2g\gamma}}}
\def\ag    {\ensuremath{\mathcal{A}_{\gamma}}}
\def\hq    {\ensuremath{\mathcal{H}_{q}}}
\def\lr    {\ensuremath{\left(}}
\def\rr    {\ensuremath{\right)}}
\def\la    {\ensuremath{\left|}}
\def\ra    {\ensuremath{\right|}}
\def\lq    {\ensuremath{\left[}}
\def\rq    {\ensuremath{\right]}}

\newcommand{\be}{\begin{eqnarray}}
\newcommand{\en}{\end{eqnarray}}
\newcommand{\nen}{\nonumber\end{eqnarray}}
\newcommand{\no}{\nonumber}

\title{%
A new \gp\ violating amplitude in the \jpsi\  decay?%
}
\articletype{} 
\author{%
\scalebox{.91}{R.~Baldini~Ferroli\inst{1}, 
F.~De~Mori\inst{2,3},
M.~Destefanis\inst{2,3},
M.~Maggiora\inst{2,3},
S.~Pacetti\inst{4}\email{simone.pacetti@pg.infn.it},
L.~Yan\inst{3,5},
M.~Bertani\inst{1},}
\\
\scalebox{.91}{%
A.~Calcaterra\inst{1}, 
G.~Felici\inst{1}, 
P.~Patteri\inst{1}, 
Y.~D.~Wang\inst{1,6},
A.~Zallo\inst{1},}
\\
\scalebox{.91}{D.~Bettoni\inst{7},
G.~Cibinetto\inst{7},
R.~Farinelli\inst{7},    
E.~Fioravanti\inst{7},
I.~Garzia\inst{7},
G.~Mezzadri\inst{7},
V.~Santoro\inst{7},    
M.~Savri\'e\inst{8},}
\\
\scalebox{.91}{F.~Bianchi\inst{2}, 
M.~Greco\inst{2},
S.~Marcello\inst{2},
S.~Spataro\inst{2},}
\\
\scalebox{.91}{C.~M.~Carloni Calame\inst{9},
G.~Montagna\inst{9,10},
O.~Nicrosini\inst{9},
F.~Piccinini\inst{9}}
}

\shortauthor{R. Baldini Ferroli {\it et al.}}

\institute{%
\inst{$^{1}$}\hspace{-2mm}Laboratori Nazionali dell'INFN di Frascati, Frascati, Italy  
\inst{$^{2}$}\hspace{-2mm}Dipartimento di Fisica, Universit\`a di Torino, Torino, Italy
\inst{$^{3}$}\hspace{-2mm}INFN Sezione di Torino, Torino, Italy
\inst{$^{4}$}\hspace{-2mm}Dipartimento di Fisica e Geologia, Universit\`a degli Studi di Perugia and INFN Sezione di Perugia, Perugia, Italy
\inst{$^{5}$}\hspace{-2mm}University of Science and Technology, Hanui, Hefei, R.P.C. 
\inst{$^{6}$}\hspace{-2mm}Helmholtz Institute, Mainz, Germany 
\inst{$^{7}$}\hspace{-2mm}INFN Sezione di Ferrara, Ferrara, Italy
\inst{$^{8}$}\hspace{-2mm}Dipartimento di Fisica e Scienze della Terra, Universit\`a di Ferrara, Ferrara, Italy  
\inst{$^{9}$}\hspace{-2mm}INFN Sezione di Pavia, Pavia, Italy   
\inst{\hspace{-1mm}$^{10}$}\hspace{-2mm}Dipartimento di Fisica, Universit\`a di Pavia, Pavia, Italy      
}

\abstract{The \jpsi\ meson has negative \gp\ so that, in the limit of isospin conservation, its decay into \pipi\ should be purely electromagnetic. However, the measured branching fraction $\B(\jpsi\to\pipi)$ exceeds by more than 3.9 standard deviations the expectation computed according to BaBar data on the $\ee\to\pipi$ cross section. The possibility that the two-gluon plus one-photon decay mechanism is not suppressed by \gp\ conservation is discussed, even by considering other multi-pion decay channels. As also obtained by phenomenological computation, such a decay mechanism could be responsible for the observed discrepancy. Finally, we notice that the BESIII experiment, having the potential to perform an accurate measurement of the $\ee\to\pipi$ cross section in the 3 GeV energy region, can definitely prove or disprove this strong \gp-violating mechanism by confirming or confuting the BaBar data.
}


\begin{document}
\maketitle
%
%
%
\section{Introduction}
\label{sec:intro}
The \jpsi\ meson as all the isoscalar vector mesons, having total angular momentum $J=1$, negative $C$-parity, $C=-1$, and isospin zero, $I=0$, posses a well defined \gp, \ie, $G=-1$. Indeed, particles that are eigenstates of the charge conjugation with eigenvalue $C$, are also eigenstates of \gp\ with eigenvalue $G=C\,(-1)^I$, where $I$ is the isospin. 
\\
\gp\ is particularly useful because it is well defined also  
for those particles, which are not $C$-parity eigenstates, as those belonging to an isospin multiplets, that have all the same value of $G$. Moreover, being a multiplicative quantum number, states containing particles, eigenstates of \gp, are themselves eigenstate of \gp\ with eigenvalue equal to the product of those of each particle. A state with $n$ pions and no other particles has total \gp, $G_{n\pi}=\lr G_\pi\rr^n=(-1)^n$, since each pion, belonging to the same isospin multiplet, has the same \gp. \ie, $G_\pi=-1$.
\\
The strong interaction conserves \gp, so that $G$ is a good quantum number in QCD, on the contrary, the \EM\ interaction can violate the isospin conservation and hence \gp.
\section{\jpsi\ decay amplitudes}
\label{sec:decay-amps}
\begin{figure}[h!]
\includegraphics[width=.9\textwidth]{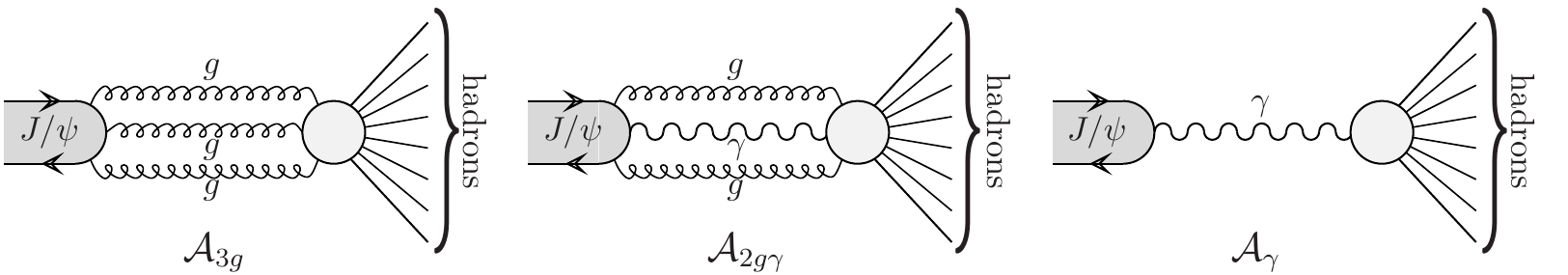}
\caption{Feynman diagrams of the three main contributions to the amplitude of the decay $\jpsi\to$ hadrons.}
\label{fig:amp}
\end{figure}

The amplitude for the decay $\jpsi\to\hq$, where \hq\ represents a final state containing only light hadrons, is usually parametrized as the sum of the three main contributions: \aggg, \agg\ and \ag, whose Feynman diagrams are shown in Fig.~\ref{fig:amp}.
%
%
\vspace{-3mm}\\
 In general the amplitude $\mathcal{A}_{\mathcal{I}}$ describes the decay chain $\jpsi\to \mathcal{I}\to \hq$, \ie, the \jpsi\ decay mediated by the virtual state $\mathcal{I}$, that could be: three gluons, $\mathcal{I}=3g$, two gluons plus one photon $\mathcal{I}=2g+\gamma$, and a single photon, $\mathcal{I}=\gamma$. The branching fractions for these \jpsi\ decays, except for $\mathcal{I}=\gamma$, for which the one into the on-shell $\mu^+\mu^-$ final state is reported, are
\be
\B(\jpsi\to 3g)
\!\!\!&=&\!\!\!
\frac{\la\aggg\ra^2\cdot{\rm PS}_{3g}}{\gjpsi}= \frac{40(\pi^2-9)}{81\,\gjpsi}
\,\alpha_s^3(\mjpsi) \,\frac{\left|\Psi(0)\right|^2}{m_c^2}\left(1+4.9\frac{\alpha_s(\mjpsi)}{\pi}\right)\,;
\label{eq:3g-rate}\\
\B(\jpsi\to 2g+\gamma)\!\!\!&=&\!\!\!
\frac{\la\agg\ra^2\cdot{\rm PS}_{2g\gamma}}{\gjpsi}= \frac{128(\pi^2-9)}{81 \,\gjpsi}
\,\alpha_s^2(\mjpsi)\, \alpha\,\frac{\left|\Psi(0)\right|^2}{m_c^2}\left(1-0.9\frac{\alpha_s(\mjpsi)}{\pi}\right)\,;
\label{eq:2g-rate}\\
\B(\jpsi\to \mu^+\mu^-)
\!\!\!&=&\!\!\!
\frac{\left|\ag\,J_{\mu^+\mu^-}\right|^2\cdot{\rm PS}_{\mu^+\mu_-}}{\gjpsi} = \frac{64 \pi}{9\,\gjpsi}
\,\alpha^2\,\frac{\left|\Psi(0)\right|^2}{\mjpsi^2}\left(1-\frac{16}{3}\,\frac{\alpha_s(\mjpsi)}{\pi}\right)\,,
\label{eq:g-rate}
\en
where PS$_{f}$ is the phase space for the final state $f$, $m_c$ is the mass of the charm quark, $\Psi(r)$ is the $c\overline c$ wave function and the quantities in parentheses are the first-order QCD corrections at the \jpsi\ mass. Equations~(\ref{eq:3g-rate}) and~(\ref{eq:2g-rate}) represent the branching fractions for the decays of the \jpsi\ into the intermediate states $3g$ and $2g+\gamma$ considered as on-shell. The decay mode of Eq.~\eqref{eq:2g-rate} is usually considered negligible~\cite{asy} with respect to the purely electromagnetic one of Eq.~\eqref{eq:g-rate} and it has been ignored so far. This assumption will be reconsidered later on.
In the Eq.~(\ref{eq:g-rate})  the amplitude \ag\ is
contracted with the point-like \EM\ current $J_{\mu^+\mu^-}$.
The branching fraction  $\B(\jpsi\to \mu^+\mu^-)$
 can be related to that of the one-photon exchange decay of \jpsi\ into a hadronic final state, \hq, $\B_\gamma(\jpsi\to\hq)$, by considering the corresponding off-peak (evaluated at $\sqrt{q^2}=3$ GeV) total cross section, as 
\be
\B_\gamma(\jpsi\to\hq)=\B(\jpsi\to\mu^+\mu^-)\left.\frac{\sigma(\ee\to\hq)}{\sigma(\ee\to\mu^+\mu^-)}\right|_{\sqrt{q^2}=3\, {\rm GeV}}<\B(\jpsi\to\mu^+\mu^-) \,R_{\rm had}(3\,{\rm GeV})\,,
\label{eq:1gamma-hq}
\en
where $R_{\rm had}$ is the ratio of the hadronic to the muon cross section in \ee\ collisions and it is $R_{\rm had}(3\,{\rm GeV})\simeq 2.5$~\cite{pdg}. Such inequality is saturated once the sum over all possible final states is considered, so that
\be
\B_\gamma(\jpsi\to{\rm hadrons}) \equiv
\sum_{\hq}\B_\gamma(\jpsi\to\hq)=
\B(\jpsi\to\mu^+\mu^-)\,R_{\rm had}(3\,{\rm GeV})
\simeq 2.5\,\B(\jpsi\to\mu^+\mu^-)\,.
\nen
In the case of a hadronic final state with negative \gp, as those containing only an odd number of pions, the strong amplitude \aggg\ is the dominant one. 
Moreover, by using the value $\alpha_s(\mjpsi)=0.135\pm0.015$, as extracted from the data on the ratio $\B(\jpsi\to 3g)/\B(\jpsi\to 2g+\gamma)$~\cite{pdg} and eqs.~(\ref{eq:3g-rate})-(\ref{eq:g-rate}), the following ratios of branchings can be obtained
\be
\frac{\B(\jpsi\to 3g)}{\B(\jpsi\to 2g+\gamma)}\!\!\!&=&\!\!\!
\frac{5}{16}\,\frac{\alpha_s(\mjpsi)}{\alpha}\,\frac{\pi+4.9\,\alpha_s(\mjpsi)}{\pi-0.9\,\alpha_s(\mjpsi)}=7.3\pm 0.9\,,
\no\\
&&\label{eq:numbers}\\
\frac{\B(\jpsi\to 3g)}{\B(\jpsi\to \mu^+\mu^-)}
\!\!\!&=&\!\!\!
\frac{5(\pi^2-9)}{72\pi}\,
\frac{\mjpsi^2}{m_c^2}\,
\frac{\alpha_s^3(\mjpsi)}{\alpha^2}\,\frac{\pi+4.9\,\alpha_s(\mjpsi)}{\pi-16\,\alpha_s(\mjpsi)/3}=8\pm 3
\,.
\nen 
Of particular interest are the decays of \jpsi\ into final states with positive \gp, $G=+1$, as for instance those consisting in an even number of pions. Indeed, since the strong interaction conserves \gp, the tree-gluon contribution, \aggg, is suppressed and such decays proceed mainly through the intermediate states $\gamma$ and $2g+\gamma$, that, due to the presence of the photon, can violate the isospin conservation and hence \gp. Let us stress again that the $2g+\gamma$ contribution has been considered negligible with respect to the single-photon one and therefore ignored so far.

\section{Even multi-pion final states}
\label{sec:even-pi}
As already discussed in Sec.~\ref{sec:intro}, multi-pion final states, having well defined \gp, represent useful and clean channels to test different models to parametrize the decay amplitudes and hence hypotheses about the dynamical mechanisms that rule the decay.
\\
In particular, amplitudes of \jpsi\ decays into even numbers of pions, \ie, final states with $G=+1$, are assumed to be dominated by
\ag, because \gp-conservation does not allow pure gluonic intermediate states.
\\
Some \gp-violation decay, not related to an \EM\ contribution, has been observed, being interpreted as due to \gp-violation in the produced mesons, like in the case of $\rho-\omega$ or $f_{0}- a_{0}$ mixing.
\begin{figure}[h]
\includegraphics[width=\textwidth]{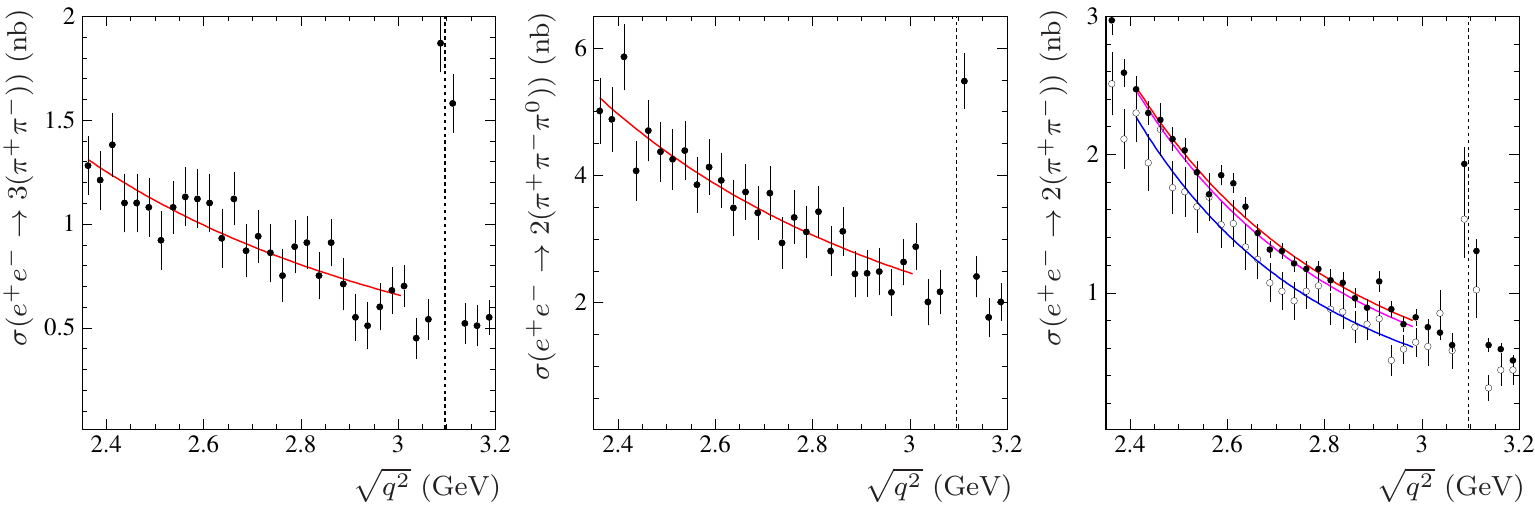}
\vspace{-3mm}
\caption{%
Data and fit on the cross sections:
$3(\pipi)$, left panel;
$2(\pipi\piz)$~\cite{babar-6pi}, central panel; 
$2(\pipi)$ from Ref.~\cite{babar-4pi-05}, empty circles, and Ref.~\cite{babar-4pi}, solid circles, right panel. In this case,
dressed data, \ie, not corrected by the vacuum polarization effects, have been considered. The fits, performed in the region $2.4\,{\rm GeV}\le \sqrt{q^2}\le 3$ GeV, are show as colored curves. In the four-pion case, right panel, two sets of data and three fits have been considered: 2012 data, red, upper curve, 2005 data, blue, lower curve, together 2005 and 2012 data, magenta, middle curve. The vertical dashed line indicates the \jpsi\ mass.}
\label{fig:even-pi}
\end{figure}\\
Figures~\ref{fig:even-pi} show cross section data and fit\footnote{\label{fn:1}Such a value has been obtained by fitting the cross section in the energy range $2.4$ GeV $\le\sqrt{q^2}\le 3$ GeV, with the power law: 
$\sigma_{\rm fit}(q^2;P_1,P_2,P_3)=P_1\,\lq(P_2^2+\lr 3\,{\rm GeV}\rr^2)/(P_2^2+q^2)\rq^{P_3}$,
where $P_1$, $P_2$ and $P_3$ are free parameters. In particular $P_1$ represents the cross section value at $\sqrt{q^2}=3$ GeV.} in the 3 GeV energy region, in case of $3(\pipi)$, $2(\pipi\piz)$ and $2(\pipi)$~\cite{babar-6pi} final states, that have been used to extract the cross section values reported in the first three rows of Table~\ref{tab:1}.
\begin{table}[h!]
\begin{center}
\renewcommand{\arraystretch}{1.3}
\begin{tabular}{c|l|c|r|r}
Decaying&
\multicolumn{1}{c|}{\multirow{2}{*}{$n\pi$ channel}} & 
$\sigma(\ee\to 2n\,\pi)$ (nb)  & 
\multicolumn{1}{c|}{\multirow{2}{*}{$\B_\gamma(\jpsi\to 2n\,\pi)$}} & 
\multicolumn{1}{c}{\multirow{2}{*}{$\B_{\rm PDG}(\jpsi\to 2n\,\pi)$}} 
\\
particle &  & at $\sqrt{q^2}=3$ GeV&&\\
\hline
\multirow{4}{*}{\jpsi}&
3(\pipi) & $0.64\pm 0.04$ & $(4.1\pm0.3)\times 10^{-3}$ & $(4.3\pm 0.4)\times 10^{-3}$\\
\cline{2-5}
&2(\pip\pim\piz) & $2.47\pm0.13$ & $(1.52\pm 0.08)\times 10^{-2}$ & $(1.62\pm 0.21)\times 10^{-2}$ \\
\cline{2-5}
&2(\pipi) & $0.73 \pm 0.02$ & $(4.50\pm0.13)\times 10^{-3}$ & $(3.57\pm0.30)\times 10^{-3}$\\
\cline{2-5}
&\pipi & $ (9\pm 3 )\times 10^{-3}$ & $(5.6\pm 1.9)\times 10^{-5}$ & $(1.47\pm0.14)\times 10^{-4}$\\
\hline
\hline
\multirow{3}{*}{$\psi(2S)$}&
\multirow{3}{*}{\pipi} &
$\sigma(\ee\to\pipi)$ (nb) &
 \multirow{3}{*}{$(2.6\pm 1.0)\times 10^{-6}$} &
 \multirow{3}{*}{$(7.8\pm 2.6)\times 10^{-6}$}
 \\
&& extrapolated at $\sqrt{q^2}=M_{\psi(2S)}$ &&\\
\cline{3-3}
&& $(2.4\pm0.8)\times 10^{-3}$&&\\
\end{tabular}
\caption{\label{tab:1}The cross section values (third column) have been obtained, as described in the text, by fitting or extrapolating  the data, which are from Ref.~\cite{babar-6pi} for the six pions, Ref.~\cite{babar-4pi,babar-4pi-05} for the
four pions, and Ref.~\cite{babar-2pi} for the two pions.
The values of the last column are from Ref.~\cite{pdg}. The last row has been inserted to highlight a similar \gp-violation phenomenon also for the $\psi(2S)$.}
\end{center}
\end{table}
\begin{figure}[h!]
\includegraphics[width=80mm]{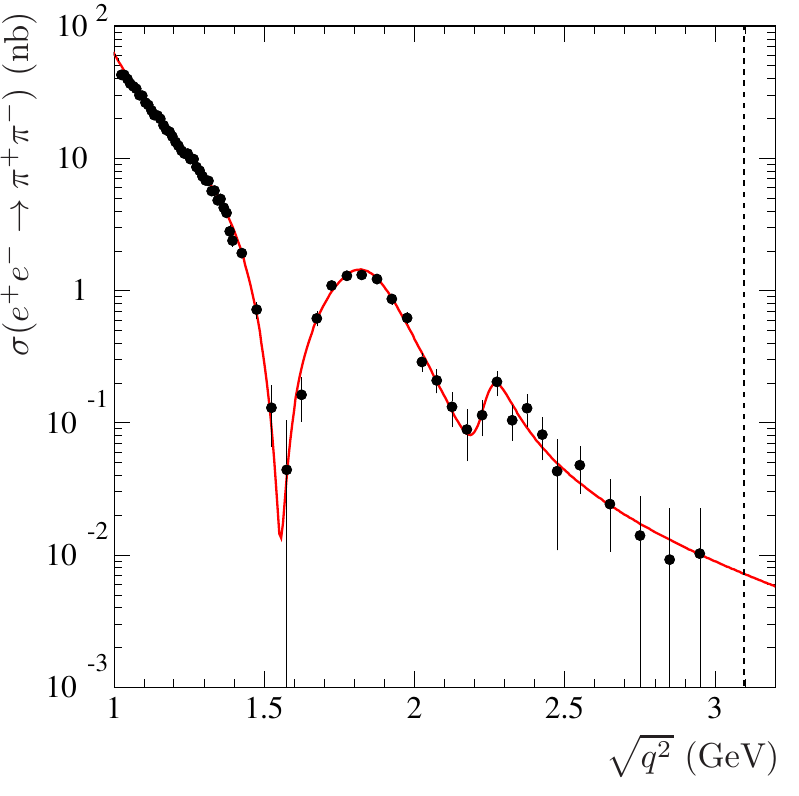}
\vspace{-5mm}
\caption{\label{fig:2pi}Data and fit on the $\pipi$ cross section~\cite{babar-2pi}. The vertical dashed line indicates the \jpsi\ mass.}
\end{figure}\vspace{0mm}\\
Concerning the \pipi\ cross section, the only set of data that reaches $\sqrt{q^2}=$ 3 GeV is the one collected by the BaBar collaboration in 2006~\cite{babar-2pi} by means of the initial state radiation techniques (ISR). However, because of
the large errors and the presence of structures nearby, the "local" fitting procedure, used in the previous cases, is unable to give reliable results. To avoid this limitation the fitting procedure of Ref.~\cite{babar-2pi} has been exploited.
The fit function, based on the Gounaris-Sakurai model~\cite{sakurai}, is shown in Fig.~\ref{fig:2pi} superimposed to the data. 
The one-photon amplitude appears as dominant, \ie, 
$\B_\gamma\simeq \B_{\rm PDG}$, in
all the multi-pion \jpsi\ decays reported in Table~\ref{tab:1},
 with the exception of the \pipi\ channel. Indeed, in this case, at most only one-half of the observed rate can be explained by 
 the contribution of \ag. The discrepancy reaches 3.9 standard deviations. 
 \\
 The BESIII experiment can measure the \pipi\ cross section in this energy region with high precision, having collected a large luminosity close to the \jpsi\ peak and more than a billion of \jpsi\ decay events. 
 \\
 If the discrepancy, observed by BaBar, would be confirmed, an additional \gp-violating decay amplitude should be considered. Such a further amplitude might strongly affect processes with branching ratios at the level of $10^{-4}$, as well as processes with branching ratios at the level of $10^{-3}$, because of the interference among the amplitudes. 
\\ 
It might be, at least in the \pipi\ case, that the one-photon amplitude does not dominate over the other \gp-violating $2g+\gamma$ contribution, that indeed should be of the same order as $\ag$, not foreseen by previous estimates~\cite{prev-estimations}. Unfortunately, it is quite difficult to compute such an amplitude in the framework of QCD, even exploiting the formulae of eqs.~(\ref{eq:3g-rate}) and~(\ref{eq:2g-rate}). Information about the relative strength of the $2g+\gamma$ amplitude with respect to the others might be inferred by considering odd-multi-pion decay channels, where \gp\ is conserved. 
In case of the four-pion channel, by assuming the one-photon dominance, the decay rate is overestimated by about 25\% and the discrepancy is about three standard deviations. 
\begin{table}[h!]
\renewcommand{\arraystretch}{1.3}
\begin{tabular}{c|c | c}
 \multirow{2}{*}{Year and Ref.}&$\sigma(\ee\to 2(\pipi))$ (nb)
 & \multirow{2}{*}{$\B_\gamma(\jpsi\to 2(\pipi))$}\\
&at 3  GeV & \\
\hline
2005~\cite{babar-4pi-05} & $0.584\pm 0.034$ & $(3.6\pm 0.2)\times 10^{-3}$\\
\hline
2012~\cite{babar-4pi} & $0.771\pm 0.019$ & $(4.76\pm 0.13)\times 10^{-3}$\\
\end{tabular}
\caption{\label{tab:2005}One-photon contributions to the decay rate of \jpsi\ into $2(\pipi)$ from 2005 and 2015 BaBar data.}
\end{table}\\
However, there exist two sets of data on the cross section $\ee\to 2(\pipi)$, both of them have been collected by the BaBar collaboration, the first  in 2005~\cite{babar-4pi-05} with an integrated luminosity of 89 fb$^{-1}$ and the second, in 2012~\cite{babar-4pi}, with an integrated luminosity of 454.3 fb$^{-1}$. In the energy region around $\sqrt{q^2}=3$ GeV these two sets give different central values for the cross sections. It is evident, see the right panel of Fig.~\ref{fig:even-pi}, that the 2005 data (empty circles) are systematically below the more accurate 2012 data (solid circles). 
Table~\ref{tab:2005} reports cross sections and decay rates obtained by fitting these two sets separately (blue and red curves in the right panel of Fig.~\ref{fig:even-pi}). %
\\
It is interesting to notice that, by considering only the older data, $\B_\gamma$ and $\B_{\rm PDG}$ agree very well, by enforcing the one-photon-dominance hypothesis. On the other hand, the possibility of a 25\% discrepancy could be explained in terms of a constructive interference effect between a dominant $\ag$
and sub-dominant $\agg\simeq \ag/6$. New measurements of such a cross section in the 3 GeV-energy region would be of great value for establishing the actual strength of the \EM\ amplitude.
\\
Finally, in the last row of Table~\ref{tab:1} we also considered
the $\psi(2S)$ decay into \pipi. To estimate the \EM\ contribution, since there are no data, we extrapolate the $\ee\to\pipi$ cross section at the $\psi(2S)$ mass by using the pion form factor parametrization as obtained in Ref.~\cite{babar-2pi}. 
Even in the case of $\psi(2S)$ as in that of \jpsi, the \EM\ contribution is responsible of only about one third of the measured branching fraction.
\subsection{The \gp\ conserving channels}
\label{subsec:psipipi}
As a reference the \gp-conserving decays
$\jpsi\to\pipi\piz$ and $\jpsi\to2(\pipi)\piz$ are considered. The corresponding production cross sections in \ee\ annihilation have been measured by the BaBar Collaboration~\cite{babar-3pi,babar-5pi}, again by means of ISR, up to center of mass energies of $\sqrt{q^2}=3$~GeV and $\sqrt{q^2}=4.5$~GeV, respectively. The values of such cross sections at $\sqrt{q^2}=3$~GeV, \ie,  
\be
\sigma(\ee\to\pipi\piz)(3\,{\rm GeV})\ug 0.063\pm 0.024\,{\rm nb}\,,\no\\
\sigma(\ee\to2(\pipi)\piz)(3\,{\rm GeV})\ug 0.26\pm 0.04\,{\rm nb}\,,
\nen
are obtained by means of the fitting procedure\footnote{See foot note~\ref{fn:1}.} used in Sec.~\ref{sec:even-pi} and shown in Fig.~\ref{fig:even-pi} together with the cross section data.
\\
The \EM\ decay rates can be computed by exploiting Eq.~(\ref{eq:1gamma-hq}), as
\be
\B_\gamma(\jpsi\to\pipi\piz)\ug \B(\jpsi\to\mu^+\mu^-)\left.\frac{\sigma(\ee\to\pipi\piz)}{\sigma(\ee\to\mu^+\mu^-)}\right|_{\sqrt{q^2}=3\,{\rm GeV}}=(3.9\pm 1.5)\times 10^{-4}\,,
\label{eq:b-gamma-3pi}\\
\B_\gamma(\jpsi\to2(\pipi)\piz)
\ug \B(\jpsi\to\mu^+\mu^-)\left.\frac{\sigma(\ee\to2(\pipi)\piz)}{\sigma(\ee\to\mu^+\mu^-)}\right|_{\sqrt{q^2}=3\,{\rm GeV}}=(1.63\pm 0.24)\times 10^{-3}\,,
\label{eq:b-gamma-5pi}
\en
to be compared with  the PDG data
\be
 \B_{\rm PDG}(\jpsi\to\pipi\piz)\ug (2.11\pm0.07)\times 10^{-2}\,,\no\\
 \B_{\rm PDG}(\jpsi\to2(\pipi)\piz)
 \ug(4.1\pm0.5)\times 10^{-2}\,.
 \nen
Assuming that such decays are dominated by the three-gluon exchange mechanism, whose Feynman diagram is shown in the left panel of Fig.~\ref{fig:amp}, the branchings can be parametrized following Eq.~(\ref{eq:3g-rate}) as
\be
\B_{3g}(\jpsi\to 3\pi,5\pi)\ug 
\B(\jpsi\to 3g)\lq \frac{4}{3}\ajpsi \rq^3\cdot{\rm PS}_{3\pi,5\pi}
\no\\
\ug
\frac{40(\pi^2-9)}{81\,\gjpsi}
\,\alpha_s^3(\mjpsi) \,\frac{\left|\Psi(0)\right|^2}{m_c^2}\left(1+4.9\frac{\alpha_s(\mjpsi)}{\pi}\right)
\lq \frac{4}{3}\ajpsi \rq^3\cdot{\rm PS}_{3\pi,5\pi}\,,
\nen
where the factor $\lq 4\ajpsi/3 \rq^3$ accounts for the three gluon vertices in the final state, while PS$_{3\pi,5\pi}$ represents the three, five-pion phase space. In the same line of reasoning, the $2g+\gamma$ contributions, central panel of Fig.~\ref{fig:amp}, are obtained from Eq.~(\ref{eq:2g-rate}) as
\be
\B_{2g\gamma}(\jpsi\to3\pi,5\pi)\ug 
\B(\jpsi\to 2g+\gamma)
\lq \frac{4}{3}\ajpsi \rq^2\,\alpha \cdot{\rm PS}_{3\pi,5\pi}
\no\\
\ug
\frac{128(\pi^2-9)}{81 \,\gjpsi}
\,\alpha_s^2(\mjpsi)\, \alpha\,\frac{\left|\Psi(0)\right|^2}{m_c^2}\left(1-0.9\frac{\alpha_s(\mjpsi)}{\pi}\right)
\lq \frac{4}{3}\ajpsi \rq^2\,\alpha \cdot{\rm PS}_{3\pi,5\pi}\,,
\nen
where, with respect to the previous case, there is only the exchange of a gluon propagator with a photon propagator, hence    
there are two powers of \ajpsi\ and one of the \EM\ coupling constant $\alpha$, while the phase space is the same.
Using the value $\ajpsi=0.135\pm 0.015$ obtained in Sec.~\ref{sec:decay-amps}, the first ratio of Eq.~(\ref{eq:numbers}) and assuming that the PDG value is dominated by the three-gluon exchange contribution one gets
\be
\B_{2g\gamma}(\jpsi\to\pipi\piz)\ug
\B_{\rm PDG}(\jpsi\to\pipi\piz)
\frac{\B(\jpsi\to 2g+\gamma)}{\B(\jpsi\to 3g)}\,\frac{\alpha}{4\ajpsi/3}=(1.2\pm 0.2)\times 10^{-4} \,,\no\\
\B_{2g\gamma}(\jpsi\to2(\pipi)\piz)\ug
\B_{\rm PDG}(\jpsi\to2(\pipi)\piz)
\frac{\B(\jpsi\to 2g+\gamma)}{\B(\jpsi\to 3g)}\,\frac{\alpha}{4\ajpsi/3}=(2.3\pm 0.3)\times 10^{-4} \,.
\nen
\begin{figure}[h!]
\includegraphics[width=.8\textwidth]{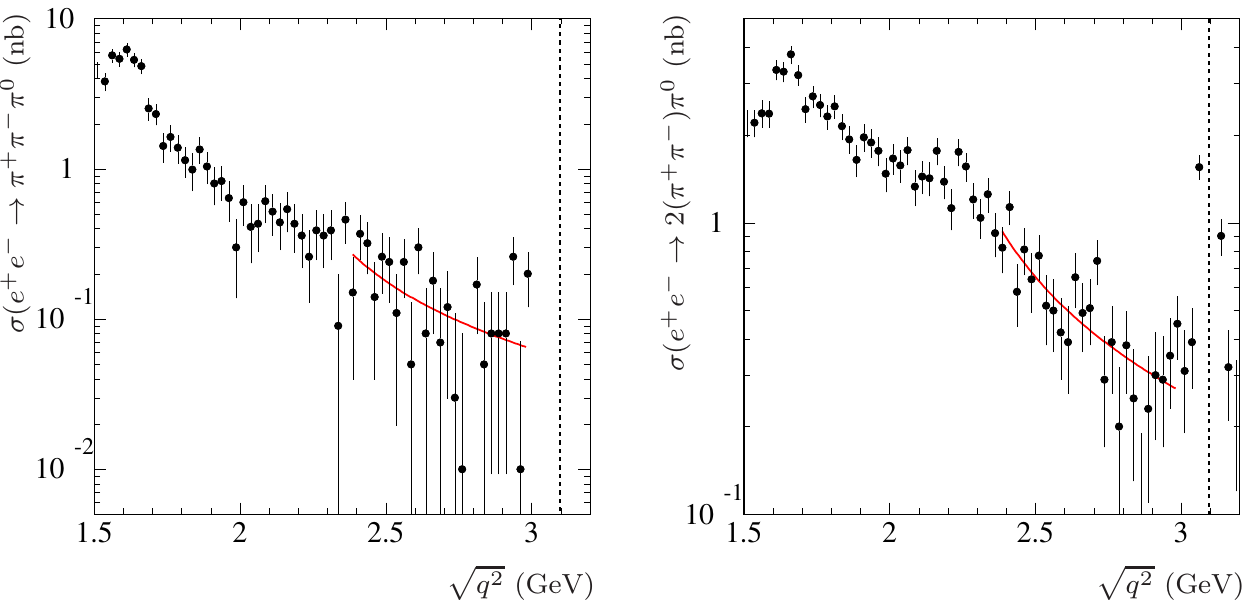}
\vspace{-2mm}
\caption{The solid points represent the data on 
$\pipi\piz$~\cite{babar-3pi} (left) and $2(\pipi)\piz$~\cite{babar-5pi} (right) cross section. The curves (red in the on-line version) are the fits, performed in the region $2.4\,{\rm GeV}\le \sqrt{q^2}\le 3$ GeV, and the vertical dashed lines indicate the \jpsi\ mass.}
\label{fig:odd-pi}
\end{figure}\vspace{0mm}\\
It is interesting to notice that, while the three-pion-$(2g+\gamma)$ rate is of the same order of $\B_\gamma(\jpsi\to\pipi\piz)$, given in Eq.~(\ref{eq:b-gamma-3pi}), the five-pion-$(2g+\gamma)$ rate
is one order of magnitude lower than the \EM\ one, Eq.~(\ref{eq:b-gamma-5pi}).
\begin{table}[h!]
\begin{center}
\renewcommand{\arraystretch}{1.3}
\begin{tabular}{l|c|r|r|r}
\multicolumn{1}{c|}{\multirow{2}{*}{$2(n\!+\!1)\,\pi$}} & 
$\sigma\lr\ee\to (2n\!+\!1)\,\pi\rr$ (nb) & 
\multicolumn{1}{c|}{\multirow{2}{*}{$\B_\gamma\lr\jpsi\to (2n\!+\!1)\,\pi\rr$}}& 
\multicolumn{1}{c|}{\multirow{2}{*}{$\B_{2g\gamma}\lr\jpsi\to (2n\!+\!1)\,\pi\rr$}} & 
\multicolumn{1}{c}{\multirow{2}{*}{$\B_{\rm PDG}\lr\jpsi\to (2n\!+\!1)\,\pi\rr$}} \\
\multicolumn{1}{c|}{channel} & at $\sqrt{q^2}=3$ GeV &&&\\
\hline
\pipi\piz & $0.063\pm 0.024$ &
$(3.9\pm 1.5)\times 10^{-4}$ & $(1.2\pm0.2)\times 10^{-4}$ & $(2.11\pm 0.07)\times 10^{-2}$\\
\hline
2(\pipi)\piz & $0.26\pm 0.04$ &
$(1.63\pm 0.24)\times 10^{-3}$ & $(2.3\pm 0.3)\times 10^{-4}$ & $(4.1\pm 0.5)\times 10^{-2}$ \\
\end{tabular}
\caption{\label{tab:2}%
Cross sections and, one-photon and two-gluon plus one-photon contributions
to the branching fractions of the \gp-conserving channels 
$\pipi\piz$ and $2(\pipi)\piz$, compared with the data from Ref.~\cite{pdg}, reported in the last column.}
\end{center}
\end{table}
\\
The different hierarchies among the contributions in these two channels and, in particular, the fact that $\B_\gamma$ and $\B_{2g\gamma}$ are of the same order in case of \pipi\piz, while $\B_\gamma\gg\B_{2g\gamma}$ in case of $2(\pipi)\piz$ is due to the values of the cross sections in \ee\ annihilation. 
The cross section decreases with the pion multiplicity faster than the decay rate, indeed (at $\sqrt{q^2}=3$ GeV)  
\be
\frac{\sigma(\ee\to \pipi\piz)}{\sigma(\ee\to 2(\pipi)\piz)}\sim \frac{1}{4}\,,
\hh
\frac{\B_{2g\gamma}\lr\jpsi\to\pipi\piz\rr}{\B_{2g\gamma}\lr\jpsi\to2(\pipi)\piz\rr}=
\frac{\B_{\rm PDG}\lr\jpsi\to\pipi\piz\rr}{\B_{\rm PDG}\lr\jpsi\to2(\pipi)\piz\rr}\sim\frac{1}{2}\,.
\nen 
In other words, the drop of the \ee\ cross section value as the pion multiplicity in the final state decreases, makes the one-photon contribution comparable to the $2g+\gamma$ one. However, the dominance of the three-gluon amplitude in the \gp-conserving channels hides this effect. On the contrary, in the \gp-violating decays of the \jpsi, where the \aggg\ amplitude is suppressed, the effect of the drop of $\B_\gamma/\B_{2g\gamma}$ as the pion multiplicity
decreases, becomes important being \agg\ and \ag\ the dominant amplitudes.
\\
In light of that, it is plausible that for the \pipi\ final state, \ie, the multi-pion channel with the lowest multiplicity, the amplitudes \agg\ and \ag\ are similar and hence by considering \ag only, as done in Sec.~\ref{sec:even-pi} and shown in Table~\ref{tab:1}, the decay rate would be underestimate. 
\\
A computation of the \agg\ contribution, made by means of a procedure based on a phenomenological description of the $2g+\gamma$ coupling, the Cutkosky rule~\cite{cut-rule} and the dispersion relations, has been made in Ref.~\cite{bal-man-pac}. The obtained value
\be
\B_{2g\gamma}(\jpsi\to\pipi)=(5.78\pm 0.45^{\rm stat}\pm 0.43^{\rm syst})\times 10^{-5}\,,
\label{eq:pheno}
\en
where the systematic error that obtained under the hypothesis of  one-photon-exchange dominance given in Table~\ref{tab:1}, i.e., $\B_\gamma(\jpsi\to\pipi)=(5.6\pm 1.9)\times 10^{-5}$. 
\section{The CLEO datum}
\label{sec:cleo}
\begin{figure}[h!]
\includegraphics[width=85mm]{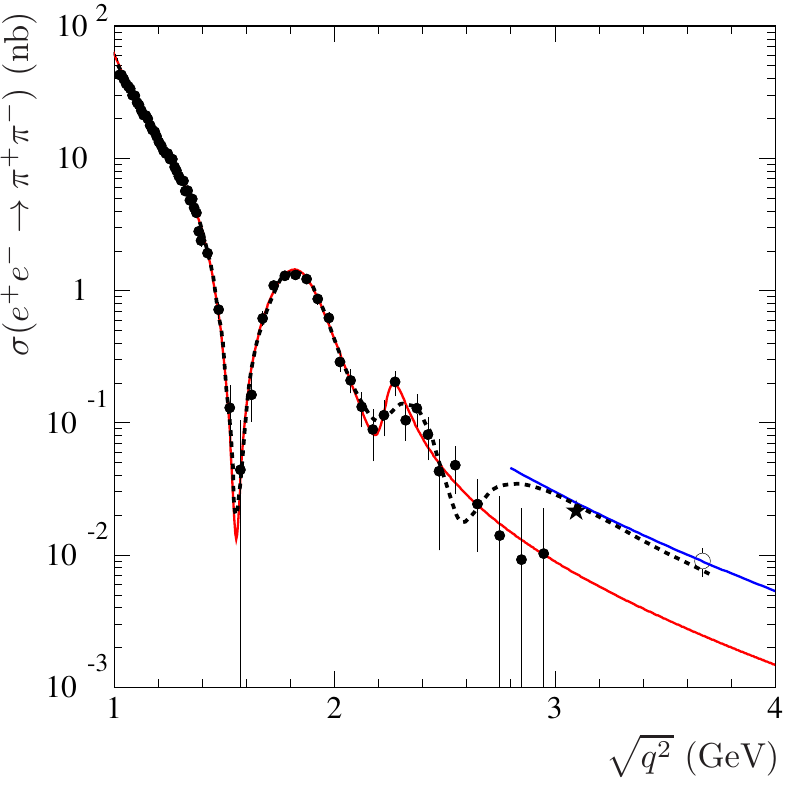}
%
\caption{\label{fig:2pi-cleo}The CELO~\cite{cleo-2pi} datum (empty circle) together with BaBar~\cite{babar-2pi} points (solid circles) and a theoretical estimate~\cite{milana} of the $\pipi$ cross section at the \jpsi\ mass (solid star). The red curve is the fit on the BaBar data, the black dashed curve is the fit of Ref.~\cite{czyz} and
 the blue curve is cross section extrapolated from the CLEO point assuming the perturbative QCD power law~\cite{asy}.}
\end{figure}

Another unexpected result is represented by the high 
$\ee\to\pipi$ cross section measured by the CLEO Collaboration~\cite{cleo-2pi} close to the $\psi(2S)$ mass, at $\sqrt{q^2}=3.671$ GeV. The cross section datum is shown (empty circle) in fig~\ref{fig:2pi-cleo}.
\vspace{0mm}\\
This result is unexpected because, following perturbative QCD~\cite{asy} (pQCD), at high $|q^2|$, the pion form factor should vanish with the power law $\lr q^2\rr^{-1}$, as a consequence, the cross section scales as
 \be
\sigma(\ee\to\pipi)\lr \sqrt{q^2}\rr\mathop{\propto}_{|q^2|\to\infty}
\lr\frac{1}{q^2}\rr^3\,.
\nen
Assuming the power-law behavior and relying on the only CLEO point, the cross section extrapolated at 3 GeV, blue curve in Fig.~\ref{fig:2pi-cleo}, is
\be
\sigma(\ee\to\pipi)(3 \,{\rm GeV})_{\rm CLEO}=\lr 0.030\pm 0.007
\rr\,{\rm nb}\,.
\nen
This value is more than three times higher than that, reported in Table~\ref{tab:1}, obtained by the extrapolation of the BaBar data, red curve in Fig.~\ref{fig:2pi-cleo}, and, through the formula of Eq.~(\ref{eq:1gamma-hq}), it gives the electromagnetic branching 
\be
\B_\gamma^{\rm CLEO}(\jpsi\to\pipi)
=\lr1.85\pm 0.43\rr\times 10^{-4}
\,,
\nen
that, being in agreement with the PDG value
$\B_{\rm PDG}(\jpsi\to\pipi)=\lr1.47\pm0.14\rr\times 10^{-4}$, confirms \gp\ 
conservation, \ie,  the one-photon-exchange dominance in the decay $\jpsi\to\pipi$.
\\
However, as can be seen in Fig.~\ref{fig:2pi-cleo}, the two extrapolations, from BaBar data to higher $q^2$'s and from the CLEO point, back, to lower $q^2$'s, are not compatible, that is, BaBar and CLEO data do not follow the pQCD behavior.
\\
There are then three possibilities:
\begin{itemize}
\item
the BaBar measurement underestimates the cross section in the region $2.3-3.0$ GeV by a factor of three;
\item
the CLEO datum overestimates the cross section at $\sqrt{q^2}=3.671$ GeV by a factor of three;
\item
the high-$q^2$ regime at which pQCD is expected to hold is still not reached, \ie, other prominent structures (strongly coupled high-mass resonances) are present and then, BaBar and CLEO data are actually compatible. 
\end{itemize}
The last possibility has been considered in Ref.~\cite{czyz}, where the authors fit all the pion form factor data, including not only the CLEO point, but also a theoretical value~\cite{milana} at the \jpsi\ mass, star symbol in Fig.~\ref{fig:2pi-cleo}, obtained from the branching ratio $\B_{\rm PDG}(\jpsi\to\pipi)$, assuming \gp\ conservation. The cross section obtained in Ref.~\cite{czyz} is shown as a black dashed curve in Fig.~\ref{fig:2pi-cleo}. The structure found at $\sqrt{q^2}\simeq 2.8$ GeV is due to the model used to fit the pion form factor data, which accounts, not only for the "visible" resonances, but also for the infinite possible $\rho$ radial excitations~\cite{dual-qcd}. Nevertheless the last three BaBar points, with $\sqrt{q^2}\ge 2.7$ GeV, are hardly described.
\\
In light of all this, an accurate measurement of the $\ee\to\pipi$ cross section in the \jpsi\ mass region appears as the only and inescapable means to clarify these items.
\section{The weird case of \opiz} 
\label{app:omega-piz}
\begin{figure}[h!]
\includegraphics[width=120mm]{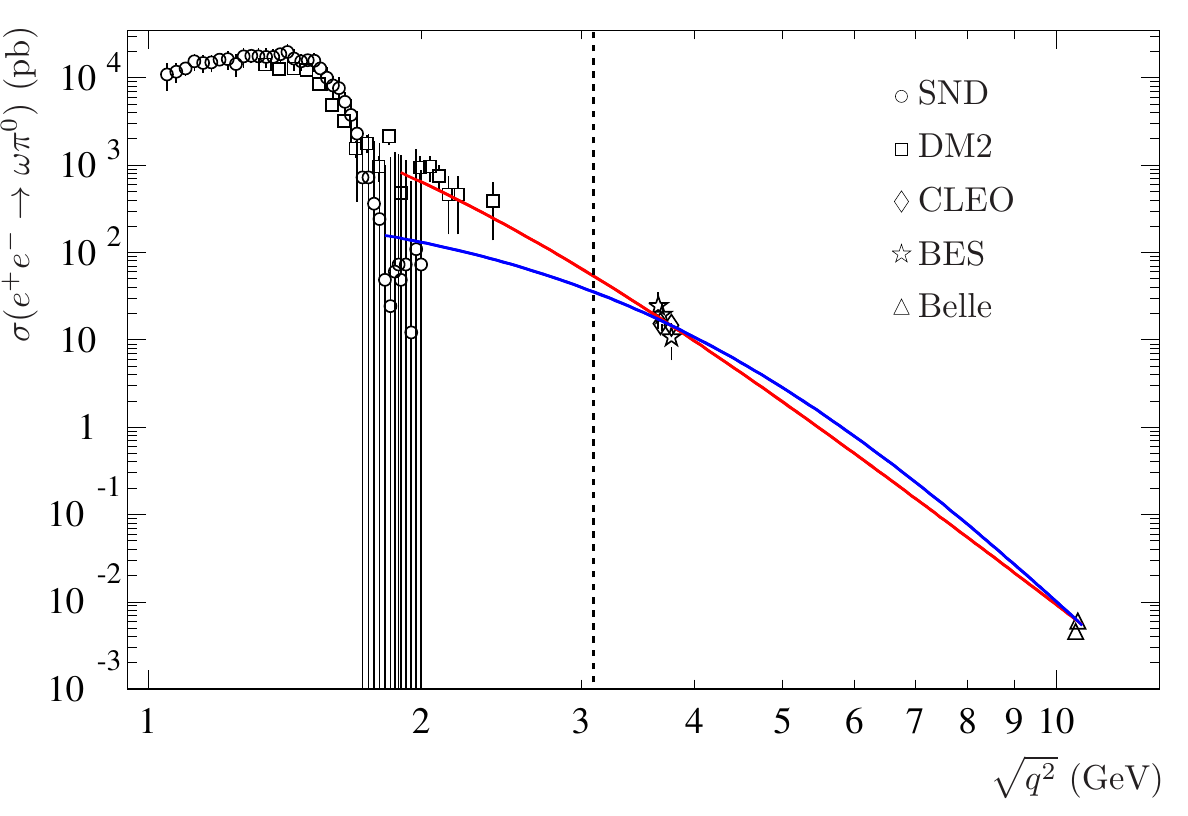}
\vspace{-5mm}\\
\caption{Data on $\sigma(\ee\to\opiz)$ from SND (empty circles)~\cite{snd}, DM2 (empty squares)~\cite{dm2}, CLEO (empty diamonds)~\cite{cleo} and Belle (empty triangles)~\cite{belle}. The red and blue curves represents the fits described in the text, and the vertical dashed line indicates the \jpsi\ mass.}
\label{fig:opi-data}
\end{figure}

The decay $\jpsi\to\opiz$, with a branching fraction $\B_{\rm PDG}(\jpsi\to\opiz)=(4.5\pm 0.5)\times 10^{-4}$~\cite{pdg}, could be another channel where \gp\ is violated. Unfortunately there are no data on the cross section $\sigma(\ee\to\opiz)$ at $\sqrt{q^2}=3$ GeV, that can be used to estimate, through Eq.~(\ref{eq:1gamma-hq}), the \EM\ contribution, $\B_{\gamma}(\jpsi\to\opiz)$. Nevertheless, data on such a cross section are available in other energy regions. In particular, as shown in Fig.~\ref{fig:opi-data}, at low $q^2$, the DM2~\cite{dm2} Collaboration collected data in the range $(1.05 \le \sqrt{q^2}\le 2.00)$ GeV, while the SND Collaboration~\cite{snd} covered the interval $(1.35 \le \sqrt{q^2}\le 2.40)$ GeV. BES~\cite{bes-opi} and CLEO~\cite{cleo}  measured $\sigma(\ee\to \opiz)$ around the $\psi(2S)$ and $\psi(3770)$ masses, collecting three and two data points, respectively, that are shown as empty stars and diamonds  in Fig.~\ref{fig:opi-data}. Finally, the Belle Collaboration~\cite{belle}, took data on the same cross section in proximity of the  $\Upsilon(4S)$ mass, the two point are reported in Fig.~\ref{fig:opi-data} as empty triangles. 
Following pQCD the expected asymptotic behavior for the cross section $\sigma(\ee\to\opiz)$ as a function of $q^2$ is~\cite{asy}
\be
\sigma(\ee\to\opiz)\lr \sqrt{q^2}\rr
\mathop{\propto}_{q^2\to\infty} 
\la F_{\opiz}(q^2)\ra^2
\mathop{\propto}_{q^2\to\infty}
\lr q^2\rr^{-4}\,.
\nen\vspace{-3mm}\\
%
%
\vspace{3mm}\\
In light of this, to obtain the value at $\sqrt{q^2}=3$ GeV, the high energy data are fitted with\footnote{See foot note~\ref{fn:1}.}
\be
\sigma_{\rm asy}(q^2;P_1,P_2,P_3)=P_1\lr\frac{P_2^2+(3\,{\rm GeV})^2}{P_2^2+q^2}\rr^{P_3}\,,
\nen
where $P_1$, $P_2$, $P_3$ are free parameters and $P_1$ represents the
desired value of the cross section. Moreover, since the high energy tails of DM2 and SND data disagree, two fits have been performed by considering at low energy either the only DM2 data with $\sqrt{q^2}\ge 1.9$ GeV, or the only SND data with $\sqrt{q^2}\ge 1.825$ GeV. These two lower limits have been chosen to have the same number of points from both DM2 and SND data set. At higher energies, in both cases, all the available data from BES, CLEO and Belle have been included. The two results, called DM2 and SND cases, are shown in Fig.~\ref{fig:opi-data} as curves (blue and red in the on-line version) superimposed to the data. The parameters and normalized $\chi^2$'s are reported in Table~\ref{tab:omega-pi0}.
\begin{table}[h!]
\renewcommand{\arraystretch}{1.5}
\begin{tabular}{l|c|c|c|c}
Case & $P_1$  (pb) & $P_2$ (GeV) & $P_3$ &  $\chi^2/$d.o.f. \\
\hline
DM2 & $66\pm 4$ &   $1.64\pm 0.13$ &  $4.07\pm 0.04$  & 2.64\\
\hline
SND & $40.0\pm 1.6$ &   $4.08\pm 0.03$ &  $5.47\pm 0.03$  & 2.24\\
\end{tabular}
\caption{\label{tab:omega-pi0}Best parameters and normalized $\chi^2$'s for the fit function describing the $\omega\piz$ cross section.}
\end{table}\\
In the DM2 case, despite the large $\chi^2$/d.o.f, the value of the $P_3$ parameter, which defines the power-law behavior, is in perfect agreement with the pQCD expectation that is, on the contrary, violated in the SND case.
Finally, the \EM\ contributions to the \jpsi\ branching fraction in the two cases, are obtained by using the values of the $P_1$ parameter, which represents $\sigma(\ee\to\opiz)(3\,{\rm GeV})$, in Eq.~(\ref{eq:1gamma-hq}),
\be
\B_\gamma(\jpsi\to\opiz)=\B(\jpsi\to\mu^+\mu^-)\left.\frac{\sigma(\ee\to\opiz)}{\sigma(\ee\to\mu^+\mu^-)}\right|_{\sqrt{q^2}=3\, {\rm GeV}}=
\left\{\begin{array}{lcl}
(4.07\pm 0.25)\times 10^{-4} && \mbox{DM2 case}\\
&&\\
(2.47\pm 0.10)\times 10^{-4} && \mbox{SND case}\\
\end{array}
\right.\,,
\nen 
to be compared with: $\B_{\rm PDG}(\jpsi\to\opiz)=(4.5\pm 0.5)\times 10^{-4}$.
\\
The BESIII experiment could definitely shed light on that issue by measuring the $\omega\pi^0$ cross section in the 3~GeV energy region.
\section{Available datasets and prospects for new measurements}
\label{sec:pi-besiii}
\begin{figure}[h!]
\includegraphics[width=110mm]{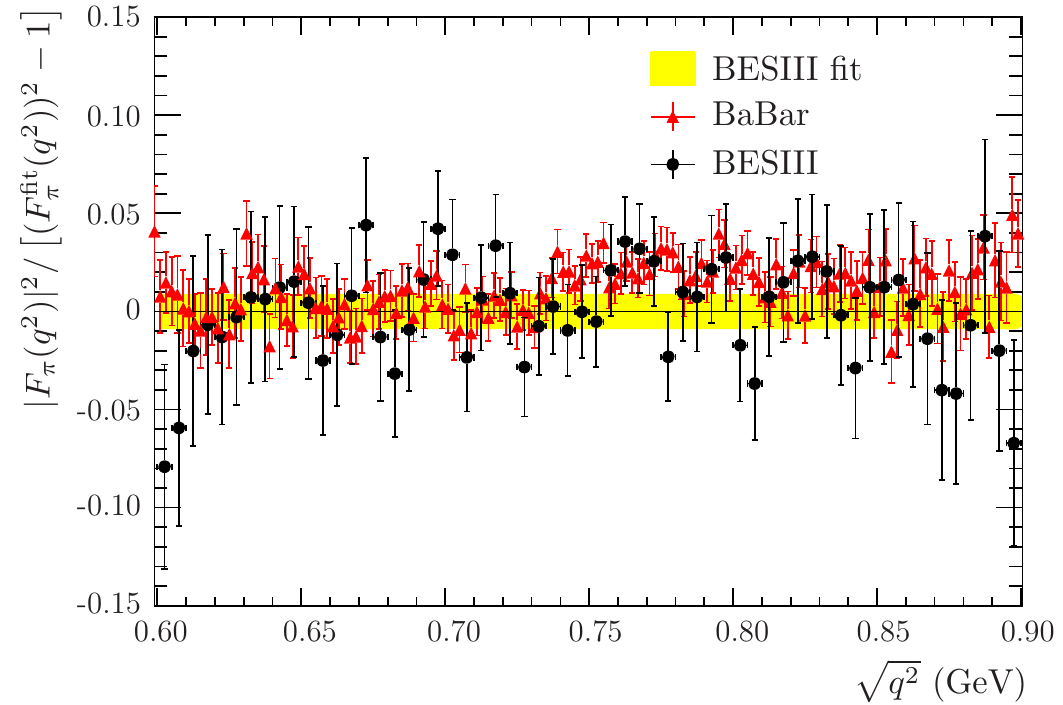}
\caption{Relative difference of the modulus squared of the pion form factor from BaBar~\cite{babar-2pi} and the BESIII fit~\cite{bes3-2pi}, the figure is from Ref.~\cite{bes3-2pi}. Statistic
and systematic uncertainties are included in the
data points. The width of the BESIII band shows
the systematic uncertainty only.}
\label{fig:bes3-babar}
\end{figure}

The CLEO Collaboration~\cite{cleo-2pi} measured the $\ee\to\pipi$ cross section at 3.671 GeV with about 20\% statistical and 15\% systematic accuracy, by collecting 20.7 pb$^{-1}$, corresponding to 26 candidate events. Pions have been identified mostly by means of the electromagnetic calorimeter. The most important background is due to the $\mu^+\mu^-$ channel and its contribution is estimated to be
they estimate that it contributes with less than 10\%. 
\\
The BESIII electromagnetic calorimeter and muon tracker~\cite{bes3} should provide similar, if not better performances. Indeed the pion shower development is expected to have a logarithmic dependence on the energy, therefore being almost the same close to the \psii\ or to the \jpsi\ mass. BESIII has collected 153 pb$^{-1}$ at 3.08 GeV and 
100 pb$^{-1}$ at 2.9 GeV, that is more than 10 times the
luminosity collected by CLEO, from which the continuum cross section was obtained. Furthermore the cross section is larger close to the  \jpsi\ mass with respect to the \psii\ and the ratio $\sigma(\ee\to\pipi)/\sigma(\ee\to\mu^+\mu^-)$ is greater too. 
\\
At the \jpsi\ mass this ratio should be the same
(or enhanced if there is the additional electromagnetic contribution,
this proposal is looking for, this ratio would be even enhanced), but the amount of events is larger and the measurement of $\B(\jpsi\to\pipi)$, having more statistics, should be easier than that of the continuum.
\\
In the BESIII experiment, the pion identification at the $\rho$ meson peak by means of ISR has been very successful~\cite{bes3-2pi}, as shown in Fig.~\ref{fig:bes3-babar}, where BESIII results are compared with the BaBar measurement~\cite{babar-2pi}.
Of course these results concern much higher $\ee\to\pipi$
cross sections, quite lower pion energies and other
kinematical constraints, and this outstanding 
achievement cannot be directly applied to our case.
However, even if not at such a high accuracy level, a precise measurement by BESIII, close to and at the \jpsi\ mass, can certainly be achieved.
%
%
\vspace{0mm}
\section{Conclusions}
\label{sec:conclu}
The \gp-violating decay $\jpsi\to\pipi$ behaves differently with respect to the other \jpsi\ decays into even-multi-pion final states. There is a non negligible disagreement, 3.9 standard deviations,  between what is expected from the measurement of the cross section close to the \jpsi\ and the measured branching ratio. 
\\
The \jpsi\ decay mechanism mediated by $2g+\gamma$, usually neglected, or better considered negligible because \gp-violating, might be responsible for this discrepancy. Indeed, it happens that for this channel the one-photon contribution is so low, that it might be of the same order of the $2g+\gamma$ one. The fact that the one-photon contribution becomes lower and lower as the pion multiplicity decreases, has been shown in Table~\ref{tab:1}, in case of even number of pions and in Table~\ref{tab:2}, in case of odd number of pions.
\\ 
In Sec.~\ref{subsec:psipipi}, it has been noticed that, for the \gp-conserving channels, the branching due to the $2g+\gamma$ intermediate state, that in this case can be estimated by exploiting its relation with the $3g$ contribution (eqs.~(\ref{eq:3g-rate}) and~(\ref{eq:2g-rate})), turns out to be comparable with $\B_\gamma$ especially in the case of $\jpsi\to\pipi\piz$.
\\
The phenomenological computation of $\B_{2g\gamma}(\jpsi\to\pipi)$ made in Ref.~\cite{bal-man-pac} corroborates the hypothesis about the softening and even the cancellation of the hierarchy between the two main contributions $\B_\gamma$ and $\B_{2g\gamma}$ in the case of lower multiplicity multi-pion final states. However, as a matter of fact, all the estimates, done until now, found the $2g+\gamma$ amplitude totally negligible with respect to the one-photon decay.  
\\
Finally, it has been shown that the BESIII experiment has the tools to repeat this measurement with high precision, to prove or disprove the discrepancy between $\B_{\gamma}(\jpsi\to\pipi)$ and $\B_{\rm PDG}(\jpsi\to\pipi)$ pointed out by the BaBar data.
\\
If confirmed, the existence of this \gp\ violating amplitude can have heavy consequences on the attempts to get the relative phase between the strong and the \EM\ \jpsi\ decay 
amplitudes, already in the case of branching ratios at the $10^{-3}$ level.
%
%
%
%
%
%
%
%
%
%

%

\begin{thebibliography}{9}
%
%
\bibitem{Kwong:1987mj}
  W.~Kwong, J.~L.~Rosner and C.~Quigg,
  Ann.\ Rev.\ Nucl.\ Part.\ Sci.\  {\bf 37} (1987) 325.
%
\bibitem{asy}
V. L. Chernyak and A. R. Zhitnitsky, JETP Lett. {\bf 25} (1977) 510;\\ 
G. P. Lepage and S. J. Brodsky, Phys. Rev. D {\bf 22} (1980) 2157;\\ 
S. J. Brodsky and G. P. Lepage, Phys. Rev. D {\bf 24}  (1981) 2848;\\
V. Chernyak, hep-ph/9906387;\\ 
V. L. Chernyak and A. R. Zhitnitsky, Phys. Rept. {\bf 112} (1984) 173;\\ 
V. V. Braguta, A. K. Likhoded, A. V. Luchinsky, Phys. Rev. D {\bf 78} (2008) 074032.
%
\bibitem{pdg} K.~A.~Olive {\it et al.} (Particle Data Group), Chin. Phys. C, {\bf 38} (2014) 090001.
%
%
\bibitem{babar-6pi}
  B.~Aubert {\it et al.}  [BaBar Collaboration],
  Phys.\ Rev.\ D {\bf 73} (2006) 052003
  [hep-ex/0602006].
%
\bibitem{babar-4pi-05}
  B.~Aubert {\it et al.}  [BaBar Collaboration],
  Phys.\ Rev.\ D {\bf 71} (2005) 052001
  [hep-ex/0502025].
%
\bibitem{babar-4pi}
  J.~P.~Lees {\it et al.}  [BaBar Collaboration],
  Phys.\ Rev.\ D {\bf 85} (2012) 112009
  [arXiv:1201.5677 [hep-ex]].
%
\bibitem{babar-2pi}
  J.~P.~Lees {\it et al.}  [BaBar Collaboration],
  Phys.\ Rev.\ D {\bf 86} (2012) 032013
  [arXiv:1205.2228 [hep-ex]].
%
\bibitem{sakurai}
G.~J.~Gounaris and J.~J.~Sakurai,
  Phys.\ Rev.\ Lett.\  {\bf 21} (1968) 244.
%
\bibitem{prev-estimations}
V.~L.~Chernyak and A.~R.~Zhitnitsky,
  Nucl.\ Phys.\ B {\bf 201} (1982) 492
   [Nucl.\ Phys.\ B {\bf 214} (1983) 547];\\
%
  Phys.\ Rept.\  {\bf 112} (1984) 173.
%
\bibitem{babar-3pi}
  B.~Aubert {\it et al.}  [BaBar Collaboration],
  Phys.\ Rev.\ D {\bf 70} (2004) 072004
  [hep-ex/0408078].
%
  \bibitem{babar-5pi}
  B.~Aubert {\it et al.}  [BaBar Collaboration],
  Phys.\ Rev.\ D {\bf 76} (2007) 092005
   [Phys.\ Rev.\ D {\bf 77} (2008) 119902]
  [arXiv:0708.2461 [hep-ex]].
%
\bibitem{cut-rule}
R.~E.~Cutkosky,
  J.\ Math.\ Phys.\  {\bf 1} (1960) 429.
%
\bibitem{bal-man-pac} R. Baldini Ferroli, A. Mangoni, S. Pacetti, "\gp\ violating amplitudes in the $\jpsi\to\pipi$ decay", to be published.
%
\bibitem{cleo-2pi}
  T.~K.~Pedlar {\it et al.} [CLEO Collaboration],
  Phys.\ Rev.\ Lett.\  {\bf 95} (2005) 261803
  [hep-ex/0510005].
%
\bibitem{milana}
  J.~Milana, S.~Nussinov and M.~G.~Olsson,
  Phys.\ Rev.\ Lett.\  {\bf 71} (1993) 2533
  [hep-ph/9307233].
%
\bibitem{czyz}
  H.~Czyz, A.~Grzelinska and J.~H.~Kuhn,
  Phys.\ Rev.\ D {\bf 81} (2010) 094014
  [arXiv:1002.0279 [hep-ph]].
%
\bibitem{dual-qcd}
C.~A.~Dominguez,
  Phys.\ Lett.\ B {\bf 512} (2001) 331
  [hep-ph/0102190].
%
\bibitem{dm2} 
D. Bisello {\it et al.}, Nucl. Phys. Proc. Suppl. {\bf 21} (1991) 111.
%
\bibitem{snd}
M.~N.~Achasov, V.~M.~Aulchenko, A.~Y.~Barnyakov, K.~I.~Beloborodov, A.~V.~Berdyugin, A.~G.~Bogdanchikov, A.~A.~Botov and T.~V.~Dimova {\it et al.},
  Phys.\ Rev.\ D {\bf 88} (2013) 5,  054013
  [arXiv:1303.5198 [hep-ex]].
  %
\bibitem{bes-opi}
M.~Ablikim {\it et al.}  [BES Collaboration],
  Phys.\ Rev.\ D {\bf 70}, 112007 (2004)
  [Phys.\ Rev.\ D {\bf 71}, 019901 (2005)]
  [hep-ex/0410031].
%
\bibitem{cleo}
G.~S.~Adams {\it et al.}  [CLEO Collaboration],
  Phys.\ Rev.\ D {\bf 73} (2006) 012002
  [hep-ex/0509011].
%
\bibitem{belle}
C.~P.~Shen {\it et al.}  [Belle Collaboration],
  Phys.\ Rev.\ D {\bf 88} (2013) 5,  052019
  [arXiv:1309.0575 [hep-ex]].
%
\bibitem{bes3}
M.~Ablikim et al. [BESIII Collaboration], Nucl. Instrum.
Meth. A {\bf 614} (2010) 345.
%
\bibitem{bes3-2pi}
  M.~Ablikim {\it et al.} [BESIII Collaboration],
  arXiv:1507.08188 [hep-ex].
%
\end{thebibliography}
\end{document}